\documentclass[twocolumn,twocolappendix]{aastex631}
\usepackage{graphicx}
\usepackage{hyperref}

\newcommand{\mum}{\ifmmode{\rm \mu m}\else{$\mu$m}\fi}

\newcommand{\figpath}{./}

\shortauthors{Sloan et al.}
\shorttitle{The Extended IRAS/LRS Atlas}

\begin{document}

\title{The Extended Atlas of Low-resolution Spectra from the
Infrared Astronomical Satellite}

\correspondingauthor{G.~C. Sloan}
\email{gcsloan@stsci.edu}

\author[0000-0003-4520-1044]{G.~C.\ Sloan}
\affiliation{Space Telescope Science Institute, 3700 San Martin Drive,
             Baltimore, MD 21218, USA}
\affiliation{Department of Physics and Astronomy, University of North
             Carolina, Chapel Hill, NC 27599-3255, USA}
\author[0000-0002-2626-7155]{Kathleen~E.\ Kraemer}
\affiliation{Institute for Scientific Research, Boston College, 140
             Commonwealth Avenue, Chestnut Hill, MA 02467, USA}
\author[0000-0002-3824-8832]{K.\ Volk}
\affiliation{Space Telescope Science Institute, 3700 San Martin Drive,
             Baltimore, MD 21218, USA}

\begin{abstract}
We present an updated atlas of spectra from the Low-Resolution
Spectrometer (LRS) on the Infrared Astronomical Satellite (IRAS), which 
took spectra from 7.67 to 22.73~\mum\ with a spectral resolving power 
($\lambda$/$\Delta$$\lambda$) of 20--60.  The updated atlas includes 
11,238 spectra, including 5425 spectra published in the original LRS 
Atlas, 5796 spectra published in three later papers, and 17 spectra 
previously available online but not published.  The updated atlas has
significantly more sources close to the Galactic plane than the original
atlas.  We have applied an improved spectral correction to remove an
artifact at 8~\mum\ in the original database.  While the IRAS mission
flew over 40 yr ago, the extended LRS atlas remains the single most 
complete database of mid-infrared spectra of nearby and bright objects 
in the Galaxy.
\end{abstract}


\section{Introduction} 

The Infrared Astronomical Satellite (IRAS) revolutionized infrared
astronomy.  Launched on 1983 January 26, IRAS was a 0.57 m infrared space 
telescope that surveyed nearly the entire sky until it exhausted its 
cryogens on 1983 November 22.  The main focal-plane array included 59 
discrete functioning detectors sensitive to radiation in four photometric 
bands:  12, 25, 60, and 100~\mum\ \citep{neu84, iras88}.

The photometry led to multiple catalogs, most notably the Point Source
Catalog (PSC) with nearly 250,000 sources.  Prior to the IRAS mission, 
the most complete infrared catalogs were the ground-based Two-Micron Sky 
Survey \citep[TMSS;][]{tmss69} with 5412 sources between declinations 
of $-$33 and $+$81, and the Air Force Geophysics Laboratory (AFGL) rocket 
survey \citep{afgl76} with 2363 sources at 4, 11, 20, and 
27~\mum.\footnote{The sky coverage of the AFGL rocket survey depended on 
the wavelength.}  These statistics show how IRAS significantly improved 
the sky and wavelength coverage of previous surveys and dramatically 
increased the number of sources known, by a factor of $\sim$50--100.  
Except for the crowded Galactic plane and a couple of gaps in coverage, 
the IRAS PSC is reasonably complete down to $\sim$0.5 Jy at 12~\mum.

In addition to the photometric data arrays, IRAS also flew with a
Low-Resolution Spectrometer (LRS) that obtained spectra from 7.67 to 
22.73~\mum.  The LRS Atlas, published soon after the mission, contained 
spectra of 5425 sources \citep{lrs86}.  To this day, this database 
represents the most complete spectral survey of the brightest objects in 
the mid-infrared sky.  Just as post-main-sequence objects dominate the 
population observed by the TMSS \citep{gra71}, they also dominate the LRS 
Atlas \citep{lrs86, hac85}.  Such is the nature of the infrared sky.  Thus, 
the LRS Atlas provides an unsurpassed opportunity to spectroscopically 
study dying stars and the dust they produce within the Galaxy.

Many revelations about evolved stars and circumstellar dust followed the 
publication of the LRS Atlas.  For example, the LRS spectra revealed the 
presence of dust components in addition to amorphous silicates in the dust 
shells around oxygen-rich stars on the asymptotic giant branch 
\citep[AGB;][]{lml86, lml90}, which led to the identification of amorphous 
alumina \citep{ona89} and crystalline alumina \citep{gla95, tak15}.  
Another example is the discovery of molecular SiO absorption at 
$\sim$8~\mum\ in the atmospheres of K giants \citep{vol89b, coh92b, 
coh92c}, which resulted in a calibration artifact in the entire LRS Atlas 
and would impact the calibration of infrared spectrometers on the Infrared 
Space Observatory \citep[ISO;][]{pri02} and the Spitzer Space Telescope 
\citep{slo15}.  These highlights are just two examples among many.  

This science resulted from the 5425 spectra in the original LRS Atlas, 
but the LRS database includes a great deal more spectral data.  As 
\cite{lrs86} explained, the full database contained 170,000 spectra of 
50,000 sources, but the published atlas included spectra for only 
$\sim$10\% of the sources.  Their most stringent criterion was that a 
source had to have two complete spectra in the multiple scans covering 
each part of the sky.  Most bright stars outside of the Galactic plane 
passed that test, and \cite{lrs86} estimated that the LRS sample was 
$\sim$90\% complete down to 28 Jy at 12~\mum\ (i.e., zero magnitude) at 
high Galactic latitudes.  Even then, they noted that ``some very bright 
red stars are conspicuously absent,'' singling out o~Cet
(Mira) and R~Leo.

Consequently, others looked to the raw database for missing sources.
\cite{vol89a} extracted the spectra of 356 additional sources brighter 
than 40 Jy at 12~\mum\ from the IRAS database (including Mira and 
R~Leo).  They concluded that no additional spectra of point sources 
brighter than that limit could be extracted.  The next 
step was to go fainter.  \cite{vol91} added 486 more sources with flux 
densities at 12~\mum\ between 20 and 40 Jy.  Finally, \cite{kwo97} 
published an extended atlas of spectra of 11,224 sources (i.e., another 
4957 previously unpublished sources).  For years, these spectra were 
accessible from a server hosted at the University of Calgary, but when 
that server went offline, the extended atlas ceased to be available to 
the general public.

Our objective is to improve the extended atlas of spectra from the LRS 
on IRAS and make those data available once again to the community.  
As new missions like the James Webb Space Telescope obtain 
spectra of individual stars in galaxies across the Local Group and, for 
bright supergiants, even from nearby groups of galaxies, the extended 
LRS atlas can serve as a powerful comparison sample of infrared spectra 
from our own Galaxy.

Section~\ref{s.atlas} describes the sample and data processing and
investigates the photometric accuracy of the spectra in the LRS
database.  Appendix~\ref{s.defsample} provides details on the
inclusion and exclusion of spectra from different atlases.  
Section~\ref{s.spec} assesses the spectral accuracy and the need for
a spectral correction, while Appendix~\ref{s.correction} explains in
more detail how that correction was determined.  
Section~\ref{s.complete} looks at the completeness of the new
extended LRS atlas, and Section~\ref{s.summary} summarizes the key 
points that users of the new database should keep in mind.
Appendix~\ref{s.class} describes the infrared spectral 
classifications that can help users navigate the database, and 
Appendix~\ref{s.access} explains how to access the new atlas and how 
the data files are organized.

\section{The extended LRS atlas \label{s.atlas}} 

\subsection{Sample \label{s.sample}} 

The sample presented here is based on the sample on the server at the
University of Calgary, with some modifications.  It includes all 5425 
sources in the LRS Atlas \citep{lrs86}.  Those sources are referred to 
here as the ``original'' sample.  It also includes all 356 spectra 
brighter than 40 Jy at 12~\mum\ added by \cite{vol89a}, all but one of 
the 486 spectra with flux densities at 12~\mum\ between 20 and 40 Jy 
added by \cite{vol91}, and 4955 of the 4957 sources added by 
\cite{kwo97}.  Together, these three groups of spectra are referred to 
as the ``supplemental'' sample.  Finally, the present sample also 
contains 17 additional spectra on the University of Calgary server but 
not included by \cite{kwo97}; these are the ``extra'' sources.  
Appendix~\ref{s.defsample} provides details of sources that 
have been rejected or renamed and the reasons why.

\subsection{Data processing \label{s.processing}} 

\begin{figure} 
\includegraphics[width=3.4in]{\figpath 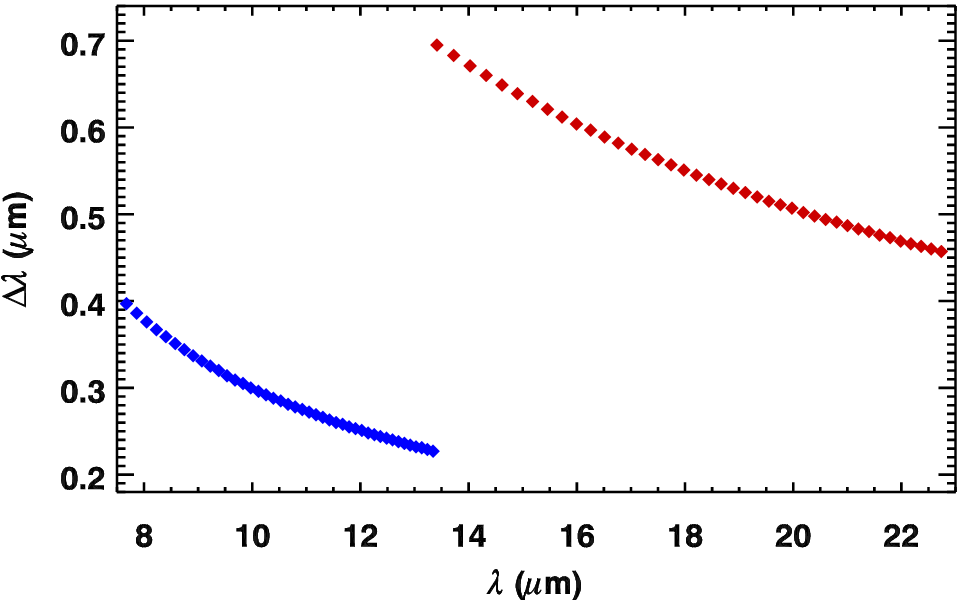} 
\caption{The resolution element ($\Delta$$\lambda$) as a function
of wavelength for the blue and red segments of the LRS at each
wavelength in the standard LRS wavelength grid.  \label{f.lrsres}}
\end{figure}

The LRS was an objective-prism spectrometer that dispersed light from 
targets in the scan direction as the spacecraft surveyed the sky.  The 
lack of a slit means that the spectral resolution is low, especially 
for extended sources.  The large beam size of the telescope and 
scanning design meant that many spectra could not be recovered in 
crowded fields.  Different detectors simultaneously obtained the blue 
and red spectral segments, covering 7.67--13.45~\mum\ and 
10.6--22.7~\mum, respectively.  Figure~\ref{f.lrsres} illustrates the 
spectral resolution of the instrument, plotting the full width at
half-maximum of the resolution element ($\Delta$$\lambda$) as a 
function of wavelength.  These values were determined from the plot of 
spectral resolving power ($R$ $\equiv$ $\lambda$/$\Delta$$\lambda$) 
versus wavelength in Chapter IX of the Explanatory Supplement 
\citep{iras88}.  The spectral resolving power varied from $\sim$20 at 
the short-wavelength end of each segment to $\sim$60 at the 
long-wavelength end.  

All of the original spectra in the LRS Atlas \citep{lrs86} were placed 
on a wavelength grid with 2 grid points per resolution element at 
7.67~\mum, smoothly increasing to 2.5 at 22.73~\mum.  The two segments 
were normalized multiplicatively using the overlap region, with the 
segment with more flagged data shifted to match the other.  Multiple 
spectra for a given target were averaged together.  The supplemental 
spectra from \cite{vol89a}, \cite{vol91}, and \cite{kwo97} and the extra 
spectra on the Calgary server followed the same processing steps as 
the original atlas.

\begin{figure} 
\includegraphics[width=3.4in]{\figpath 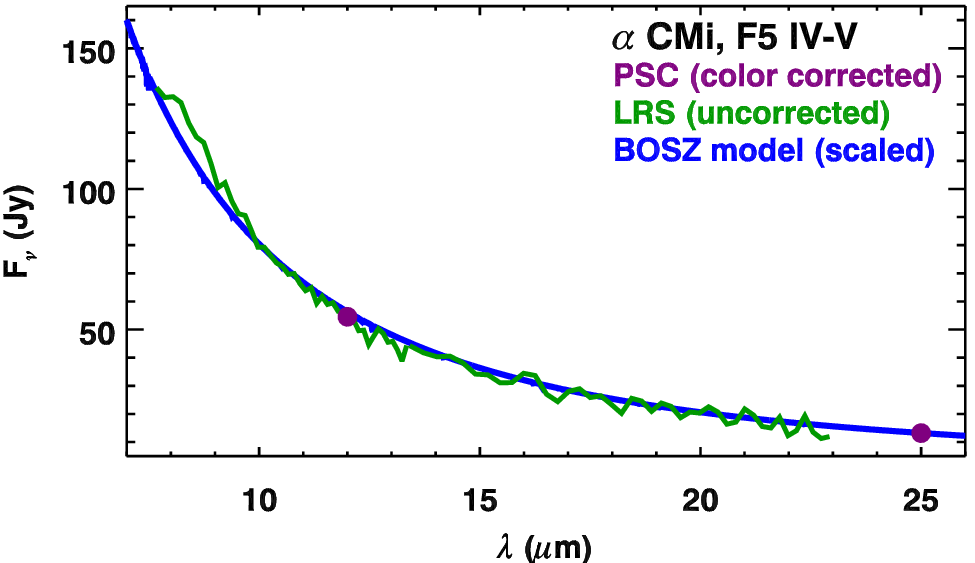} 
\caption{The uncorrected LRS spectrum of Procyon ($\alpha$ CMi), 
compared to a scaled BOSZ model and the photometry from the PSC.  The 
LRS spectrum shows the emission artifact at 8--11~\mum\ in the original 
LRS Atlas. 
\label{f.procyon}}
\end{figure}

The spectra presented here are processed in much the same way as 
before, but with two changes.  First, the spectra were trimmed, so 
that the blue segment includes 43 wavelength elements from 7.67 to 
13.34~\mum\ and the red segment includes 41 wavelength elements from 
13.41 to 22.73~\mum.  The trimming involved removing a data point at
22.92~\mum\ because of its unreliability.  Second, the spectra were 
corrected for the SiO artifact present in previously published 
spectra.  This artifact arose because the assumed truth spectra for 
K giants used as spectral standards omitted the SiO fundamental 
absorption band at 8~\mum, which was first noted by \cite{vol89b}.  
Figure~\ref{f.procyon} gives an example of the SiO artifact in the 
star $\alpha$ CMi (Procyon, F5 IV--V).  The comparison spectrum is a 
BOSZ model with $T_{\rm{eff}}$ = 6500~K, log~$g$ = 4.0, and solar 
abundances \citep{boh17, 
mes24},\footnote{\url{https://archive.stsci.edu/hlsp/bosz}} scaled to 
the LRS spectrum at 12~\mum.  To remove the spectral artifact at 
$\sim$8~\mum, we rederived the correction originally developed by 
\cite{coh92b}, as described below (Section~\ref{s.spec} and 
Appendix~\ref{s.correction}).

\subsection{Photometric accuracy of the LRS \label{s.phot}} 

\begin{figure} 
\includegraphics[width=3.4in]{\figpath 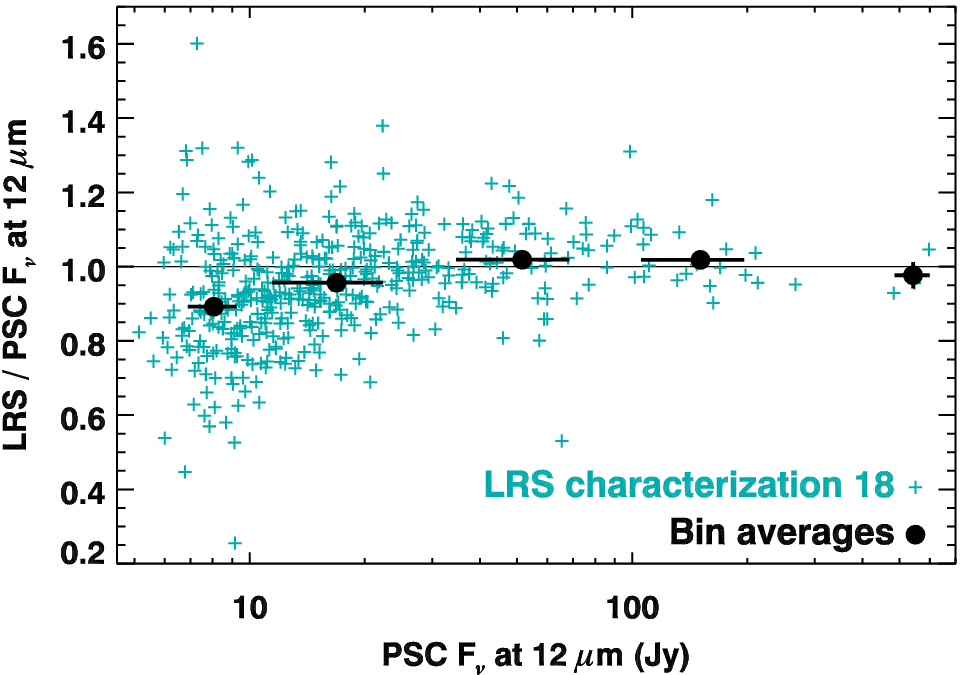} 
\caption{The ratio of the flux density at 12~\mum\ of the dust-free 
stars in the LRS Atlas from the spectra over the color-corrected
photometry from the IRAS PSC.  The large black circles give the averages 
in each half-decade bin, with horizontal error bars indicating the 
standard deviation and vertical error bars the uncertainty in the mean.  
The vertical standard deviation is apparent from the spread in the 
individual data.  \label{f.lrspsc}}
\end{figure}

\begin{deluxetable}{ccr} 
\tablecolumns{3}
\tablewidth{0pt}
\tablenum{1}
\tablecaption{Average flux densities at 12~\mum}
\label{t.lrspsc}
\tablehead{ \colhead{$\langle$$F_{\nu}$ at 12~\mum$\rangle$} & 
  \colhead{ } & \colhead{ } \\
  \colhead{from PSC (Jy)\tablenotemark{a}} & 
  \colhead{$\langle$LRS/PSC at 12~\mum$\rangle$\tablenotemark{b}} & 
  \colhead{$N$} }
\startdata
  8 $\pm$  1 & 0.89 $\pm$ 0.02 & 129 \\
 17 $\pm$  5 & 0.96 $\pm$ 0.01 & 236 \\
 52 $\pm$ 17 & 1.02 $\pm$ 0.01 &  72 \\
151 $\pm$ 45 & 1.02 $\pm$ 0.02 &  20 \\
542 $\pm$ 57 & 0.98 $\pm$ 0.04 &   3
\enddata
\tablenotetext{a}{Color-corrected; uncertainties are standard deviations
(1 $\sigma$).}
\tablenotetext{b}{Uncertainties are uncertainties in the mean.}
\end{deluxetable}

Comparing the 12~\mum\ flux densities of sources in the LRS Atlas with 
the values in the PSC reveals a discrepancy that depends on the
brightness of the source.  Figure~\ref{f.lrspsc} shows this shift for 
the dust-free stars with LRS characterizations of 18, which indicates 
that they all have a Rayleigh-Jeans tail in the LRS wavelength 
regime.\footnote{Appendix~\ref{s.lrschar} describes the LRS 
characterizations in more detail.}
Dust-free stars have a range of color corrections in the 12~\mum\
filter, from 1.41 to 1.47, depending on their effective temperature.  We 
adopted a correction of 1.45, appropriate for a 10,000 K photosphere
or a K giant with continuum opacity from the H$^-$ ion 
\citep[e.g.,][]{eng92}.  At worst, the uncertainty in color correction
leads to a 3\% uncertainty in the flux density at 12~\mum.  To determine
the 12~\mum\ flux from the LRS, the spectra were averaged from 11.8 to 
12.2~\mum\ in Rayleigh-Jeans units ($\lambda^2$$F_{\nu}$), then converted 
to Jy.

Table~\ref{t.lrspsc} gives the mean flux densities and flux ratios at
12~\mum\ for the bins plotted in Figure~\ref{f.lrspsc}.  Above 
$\sim$10~Jy the LRS/PSC ratios are consistent with 1.0, but for fainter
targets the scatter in the ratio LRS/PSC grows quite large, and the mean 
of the distribution tails off, reaching a mean discrepancy of 10\% in the 
faintest bin.  The majority of the sources with spectra from the LRS are 
in the faintest two bins.  For these fainter targets, either the PSC is 
not reliable or the LRS is not spectrophotometrically calibrated (or 
both).  The downward trend in the mean ratio LRS/PSC as the sources grow 
fainter is consistent with a slight oversubtraction of the background in 
the LRS data, but that hypothesis has not been confirmed.

\begin{deluxetable*}{lclllrrrr} 
\tablecolumns{9}
\tablewidth{0pt}
\tablenum{2}
\tablecaption{Bright red giants used as infrared standards}
\label{t.redgiants}
\tablehead{ \colhead{ } & \colhead{IRAS} & \colhead{Spectral} &
  \colhead{ } & \colhead{Used for New} & \colhead{PSC} &
  \colhead{LRS} & \colhead{Hanscom} & \colhead{LRS /} \\
  \colhead{Target} & \colhead{PSC} & \colhead{Type} & \colhead{SWS} & 
  \colhead{Correction} & \colhead{$F_{12}$ (Jy)}\tablenotemark{a} &
  \colhead{$F_{12}$ (Jy)} & \colhead{$F_{12}$ (Jy)} & \colhead{Hanscom} }
\startdata
$\beta$ Gem    & 07422$+$2808 & K0 III       & no  & yes &  85.9 &
    85.8 &  74.7 & 1.149 \\
$\alpha$ Boo   & 14133$+$1925 & K1.5 III     & yes & yes & 547.0 &
  522.6 & 458.4 & 1.140 \\
$\alpha$ TrA   & 16433$-$6856 & K2 III       & no  & no  &  99.3 &
  110.1 &  90.0 & 1.223 \\
$\alpha$ Hya   & 09251$-$0826 & K3 II--III   & no  & no  & 108.7 &
    91.9 &  89.5 & 1.027 \\
$\epsilon$ Car & 08214$-$5920 & K3 III       & no  & no  & 169.7 &
  169.2 & 158.8 & 1.065 \\
$\beta$ UMi    & 14508$+$7421 & K4 III       & yes & yes & 110.6 &
  110.9 &  98.1 & 1.130 \\
$\alpha$ Tau   & 04330$+$1624 & K5 III       & yes & yes & 482.6 &
  448.2 & 407.8 & 1.099 \\
$\gamma$ Dra   & 17554$+$5129 & K5 III       & yes & yes & 107.0 &
  118.2 &  96.4 & 1.226 \\
$\beta$ And    & 01069$+$3521 & M0 III       & yes & no  & 197.7 &
  193.2 & 171.8 & 1.125 \\
$\mu$ UMa      & 10193$+$4145 & M0 III       & yes & no  &  69.6 &
  49.9 &  64.8 & 0.770 \\
$\alpha$ Cet   & 02596$+$0353 & M1.5 III     & yes & yes & 161.9 &
  191.0 & 146.0 & 1.308 \\
$\beta$ Peg    & 23013$+$2748 & M2.5 II--III & yes & no  & 267.1 &
   254.3 & 254.1 & 1.001 \\
$\gamma$ Cru   & 12283$-$5650 & M3.5 III     & yes & yes & 596.8 &
   624.4 & 591.8 & 1.055 \\
\enddata
\tablenotetext{a}{Color-corrected by dividing by 1.45}
\end{deluxetable*}

Figure~\ref{f.lrspsc} raises questions about the photometric 
calibration of the IRAS PSC and/or the LRS.  We have also 
investigated the calibration of the LRS by comparing the LRS spectra 
of bright infrared standard stars to spectra from other telescopes, 
starting with the sample of 13 bright red giants from 
\citet[][their Table~1]{slo15}.  For each source, \cite{eng06} produced 
a carefully calibrated spectrum, in most cases based on the spectrum 
from the Short-Wavelength Spectrometer \citep[SWS;][]{lee03}, modified 
to align with the photometric calibration of the Midcourse 
Space Experiment \citep[MSX;][]{pri04}.  When SWS data were unavailable, 
infrared spectra from other sources with similar spectral types were 
used.  We will refer to these spectra based primarily on the 
SWS and MSX as the Hanscom spectra.\footnote{These spectra were produced 
at Hanscom Air Force Base and are online at
\url{https://users.physics.unc.edu/~gcsloan/library/standards}.}

We measured the 12~\mum\ flux density of each source in the LRS and
Hanscom spectra over the 11.8--12.2~\mum\ range as for the comparison
of the LRS and PSC above.  Table~\ref{t.redgiants} presents the results
and the ratios of the LRS to the Hanscom spectrum.  Excluding $\mu$~UMa, 
which is a clear outlier (and, probably not coincidentally, the faintest 
of the 13 standards), the mean ratio of LRS/Hanscom is 1.129 $\pm$ 
0.026, where the uncertainty is the uncertainty in the mean.  That is a 
significant offset, but we have not corrected the extended LRS atlas to 
remove it.  The previous comparison of the spectra from the LRS to the 
PSC revealed significant scatter in the ratio of the 12~\mum\ flux 
densities for dust-free stars and a decrease in that ratio for fainter 
targets.  These issues suggest that photometrically calibrating the 
entire database is not possible.  Consequently, users of the database 
should keep in mind the potential issues with the absolute flux 
calibration.

\section{Spectral accuracy of the LRS \label{s.spec}} 

\begin{figure} 
\includegraphics[width=3.4in]{\figpath 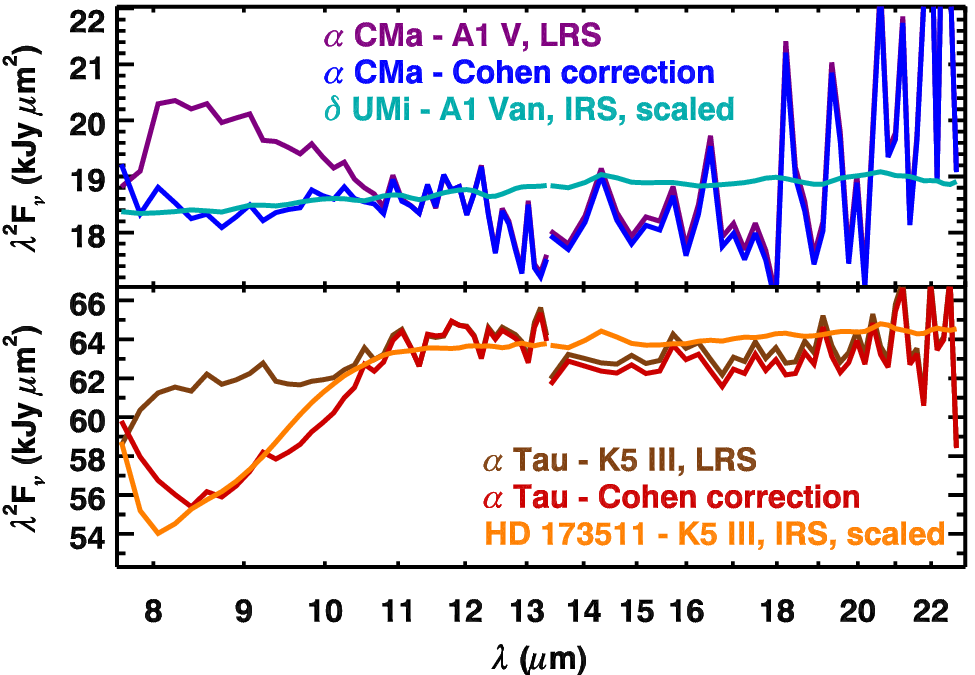}
\caption{A comparison of LRS spectra of the A dwarf $\alpha$~CMa
and the K giant $\alpha$~Tau with spectra from
the Spitzer/IRS of standard stars with similar spectral types.  The
LRS spectra are plotted with and without the spectral correction
derived by \citet[][i.e., the Cohen correction]{coh92b}.  In the 
Rayleigh-Jeans units used, stellar spectra with no absorption bands 
should be almost horizontal, with a gentle rise from the blue end, 
as $\delta$~UMi illustrates.  The spectra from the IRS are 
down-sampled to the resolution of the LRS and scaled.
\label{f.bright1}}
\end{figure}

While we have not changed the absolute photometric correction of the LRS 
data, we have applied a spectral correction from 7.67 to 10.6~\mum.  
Figure~\ref{f.procyon} illustrates the SiO artifact in that wavelength
range, but plotting the spectra in $F_{\nu}$ units hides the details.
Figure~\ref{f.bright1} examines the spectra of two well-known stars 
in Rayleigh-Jeans units, so that the Rayleigh-Jeans tail of the Planck 
function will appear as a horizontal line and any deviations are readily 
apparent.  One of the standards is the A dwarf $\alpha$~CMa (Sirius), 
which has now replaced $\alpha$~Lyr (Vega) as the primary standard star 
for all UV--optical--infrared wavelengths \citep[e.g.,][]{rie23}.  The
other is $\alpha$~Tau (Aldebaran), one of two K giants used as primary
standards in the infrared in the past (along with Arcturus, or 
$\alpha$~Boo).
Figure~\ref{f.bright1} compares their LRS spectra to spectra of standard 
stars with similar spectral types obtained with the Infrared Spectrograph 
(IRS) on the Spitzer Space Telescope \citep{slo15}.  The uncorrected LRS 
spectrum of $\alpha$~CMa shows an SiO emission artifact at 8~\mum\ when 
it should follow a Rayleigh-Jeans tail and be roughly horizontal.  The 
uncorrected LRS spectrum of $\alpha$~Tau is missing the strong SiO 
absorption band apparent in the IRS spectrum of HD~173511.

\cite{coh92b} derived a spectral correction to remove the SiO artifact
from the LRS Atlas.  Their correction, hereafter the Cohen correction,
is based on ratios of red giants to A dwarfs, most notably $\alpha$~Tau 
to $\alpha$~CMa and $\alpha$~Boo to $\alpha$~Lyr.  The published 
correction used a wavelength grid with a different break between the 
blue and red LRS segments than ours, and we have determined values for 
the six missing wavelength elements at the red end of the blue segment 
(12.81--13.34~\mum) by linear interpolation.  Applying the Cohen 
correction greatly improves the LRS spectra, but the profile of the SiO 
band in $\alpha$ Tau still does not match the profile in HD~173511.
This comparison of a spectrum from the LRS on IRAS to the IRS on 
Spitzer requires that the IRS data be down-sampled to match the much
lower resolution of the LRS; otherwise, the difference in resolution
would result in different band shapes.

\begin{figure} 
\includegraphics[width=3.4in]{\figpath 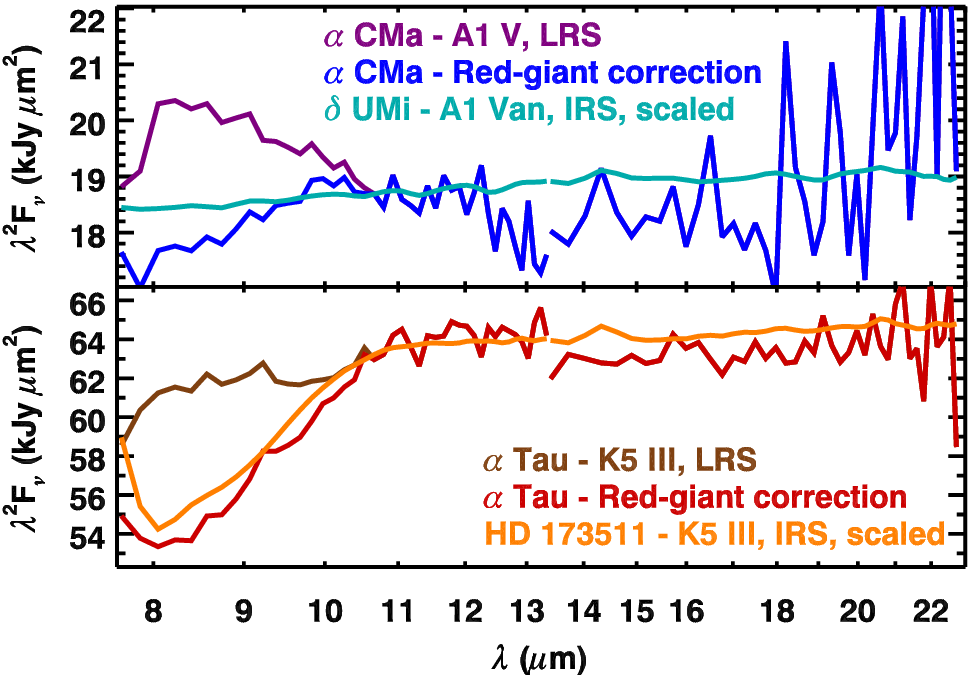}
\caption{Testing the spectral correction derived from red giants.
While the recovered shape of the SiO absorption band at 
$\sim$8--11~\mum\ has improved in $\alpha$~Tau, the shape of the 
spectrum of $\alpha$~CMa has grown worse in the same wavelength
range, especially at the blue end of the spectrum.  \label{f.bright2}}
\end{figure}

\begin{figure} 
\includegraphics[width=3.4in]{\figpath 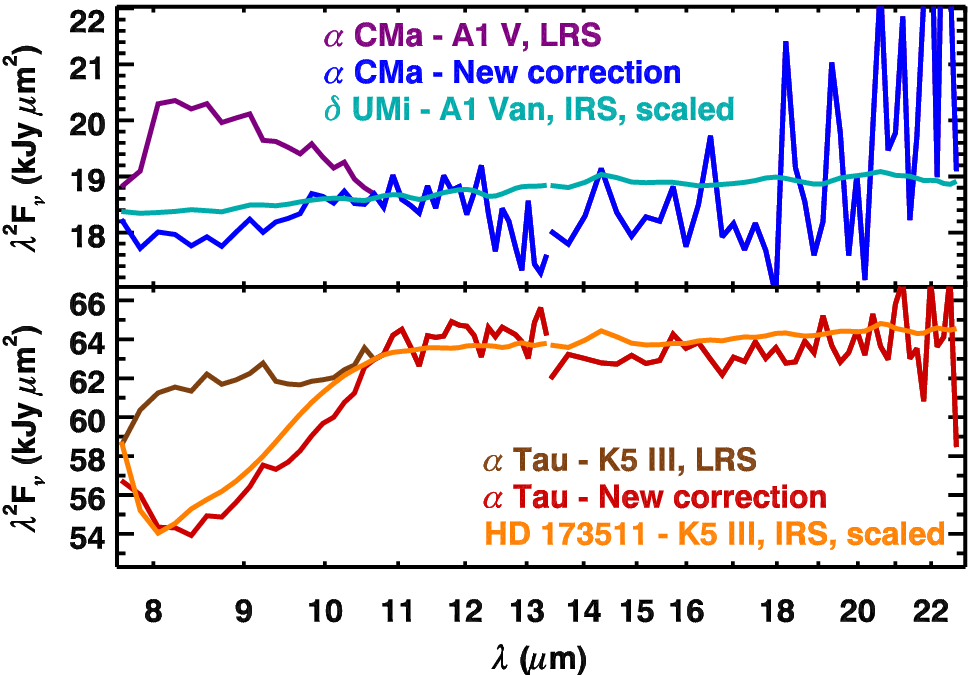}
\caption{The results of applying the new spectral correction,
which is derived from both red giants and warmer stars.  The
new correction reproduces the SiO band in a late K giant well,
and it reduces the degradation of the A dwarf at the blue
end.  \label{f.bright3}}
\end{figure}

We have improved the Cohen correction by comparing LRS spectra of 
bright red giants and warmer stars used as infrared standards.
Appendix~\ref{s.correction} describes our methodology in detail.  This 
section focuses on the results and explains the reasoning.  We 
first attempted to use just the best red giants (starting with
Table~\ref{t.redgiants}).  Figure~\ref{f.bright2} presents the results
of that effort.  While the red-giant correction improves the overall 
shape of the SiO absorption band at $\sim$8--10~\mum\ in $\alpha$~Tau, 
some differences in shape remain compared to HD~173511.  More notably, 
the spectrum of $\alpha$~CMa, which should be flat in Rayleigh-Jeans 
units, drops from $\sim$10~\mum\ to the blue end of the spectrum.

The ambiguous results from the spectral correction based on red 
giants led us to also consider using warmer stars, such as the A 
dwarfs $\alpha$~CMa and $\alpha$~Lyr.  Appendix~\ref{s.correction}
provides the details.  Combining the corrections from the red 
giants and warm stars results in a new correction which we have
adopted in place of the older Cohen correction.  Figure~\ref{f.bright3}
presents the results.  Compared to the red-giant correction, the new 
correction better reproduces the shape of the SiO band in
$\alpha$~Tau.  While it does not completely eliminate the drop 
between 7.7 and 10~\mum\ in $\alpha$~CMa, it does reduce it below 9~\mum.

The differences in the overall strength of the SiO absorption band 
in the corrected spectrum of $\alpha$~Tau compared to HD~173511 could 
easily be astrophysical in nature, as the spread in overall band 
strengths in giants of the same spectral class is well documented 
\citep{her02, eng06, slo15}.  However, the \textit{shape} of the SiO 
band is more uniform from one source to the next \citep[][see their 
Figure~13]{slo15}.  Thus, the apparent structure in the band from 8.5 
to 10.6~\mum\ in the LRS data probably indicates the limitations of 
the LRS spectra.  The deviations in $\alpha$~CMa compared to 
$\delta$~UMi tell a similar story.  The difference at $\sim$8--9~\mum\
is $\sim$2\%, which reflects the overall spectral fidelity of the
LRS database.  The deviations past 12~\mum\ are likely due to noise
in the data, because a blue star is rapidly growing fainter to longer
wavelengths.

The new spectral correction has been applied to the entire extended 
LRS atlas.  The individual spectral data files are provided with flux 
densities with and without the correction.  (Appendix~\ref{s.access} 
provides more details on the spectral data products.)

Using the two sets of standards, the red giants and the warm stars,
to generate a spectral correction should give the same answer.  The
fact that they do not points to a limitation of $\sim$2--3\% in the 
spectral fidelity of the LRS.  Thus, the LRS data are useful for
general spectral analysis but perhaps less suited to detailed 
spectral modeling.

\section{Completeness \label{s.complete}} 

\begin{figure} 
\includegraphics[width=3.4in]{\figpath 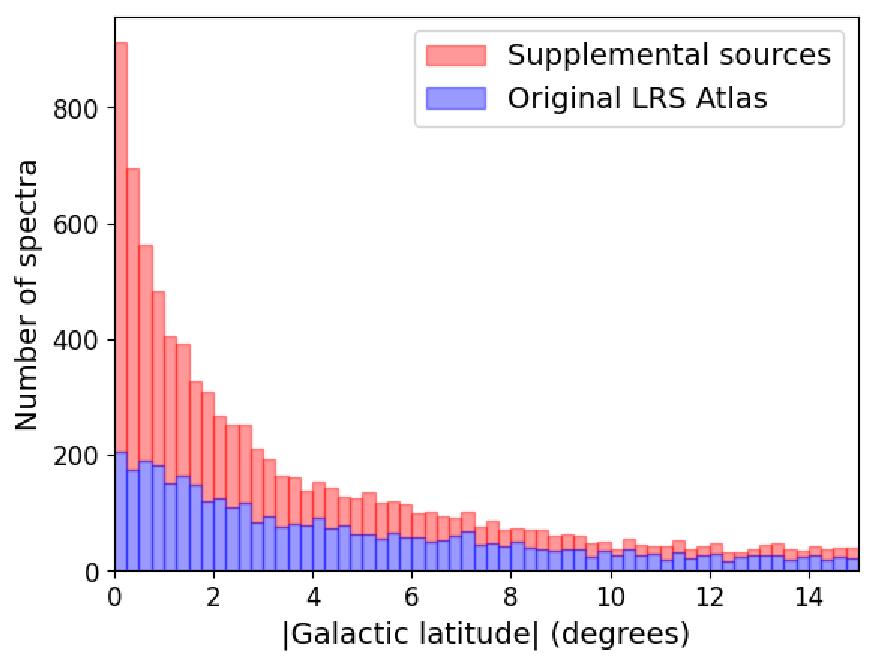}
\caption{A layered histogram showing the distribution of sources 
in the original LRS Atlas and the supplemental sources as a 
function of Galactic latitude close to the Galactic plane.  The 
bin closest to the plane has 205 original sources and 707 
supplemental and extra sources, for a total of 912 sources.  
\label{f.glathist}}
\end{figure}

Figure~\ref{f.glathist} illustrates how the original LRS Atlas is 
missing sources close to the Galactic plane.  The histogram is presented 
in layers, so that the first bin includes a total of 912 sources, 
205 in the original LRS Atlas and 707 from the supplemental and
extra sources.  The ratio of supplemental and extra sources to original 
sources in the full catalog is 1.07 (5813/5425), while the corresponding 
ratio of sources in that innermost bin is 3.45.  Within 2$\degr$ of 
the Galactic plane, the ratio is still 2.05.  Those differences should 
give some idea of how the strict selection criteria of the original LRS 
Atlas selected heavily against sources in the crowded fields near the 
Galactic plane.

\begin{deluxetable}{llrrr} 
\tablecolumns{5}
\tablewidth{0pt}
\tablenum{3}
\tablecaption{Missing sources in AKARI by LRS characterization}
\label{t.akarilrs}
\tablehead{ \colhead{LRS} & \colhead{ } & \colhead{ } & \colhead{Out} &
  \colhead{ } \\
  \colhead{Char.} & \colhead{Description} & \colhead{Missing} & \colhead{of} & 
  \colhead{Percentage} }
\startdata
0$n$ & Unusual                   &  6 &  363 &  1.7 \\
1$n$ & Blue, star                & 26 & 2238 &  1.2 \\
2$n$ & Blue, silicate emission   & 42 & 1730 &  2.4 \\
3$n$ & Blue, silicate absorption & 13 &  230 &  5.7 \\
4$n$ & C-rich dust               &  9 &  538 &  1.7 \\
5$n$ & Red, featureless          &  6 &   63 &  9.5 \\
6$n$ & Red, silicate emission    &  1 &   78 &  1.3 \\
7$n$ & Red, silicate absorption  &  7 &   67 & 10.4 \\
8$n$ & PAHs                      &  8 &   69 & 11.6 \\
9$n$ & PAHs, no lines            &  7 &   49 & 14.3
\enddata
\end{deluxetable}

\begin{deluxetable}{llrrr} 
\tablecolumns{5}
\tablewidth{0pt}
\tablenum{4}
\tablecaption{Missing sources in AKARI by VC89 class}
\label{t.akarivolk}
\tablehead{ \colhead{VC89} & \colhead{ } & \colhead{ } & \colhead{Out} & 
  \colhead{ } \\
  \colhead{Class} & \colhead{Description} & \colhead{Missing} & \colhead{of} & 
  \colhead{Percentage} }
\startdata
S & Dust-free star      &   9 & 1635 &  0.6 \\
F & Star $+$ some dust  &  37 & 1944 &  1.9 \\
E & Silicate emission   & 114 & 3617 &  3.2 \\
A & Silicate absorption &  25 &  319 &  7.8 \\
C & C-rich dust         &   8 &  715 &  1.1 \\
P & PAHs                &  55 &  315 & 17.5 \\
H & Red continuum       & 185 &  638 & 29.0 \\
U & Unusual             &  67 &  661 & 10.1 \\
L & Lines               &   0 &   31 &  0.0 \\
I & Incomplete          & 139 & 1363 & 10.2
\enddata
\end{deluxetable}

The decreasing completeness of the LRS Atlas toward the Galactic 
plane is a result of the large beam size of IRAS, which leads to 
significant uncertainties in the positions of sources.  Formally,
the IRAS resolution element at 12~\mum\ would be 5$\farcs$3 
across.  For point sources, the actual radial uncertainty has a 
Gaussian $\sigma$ = 8$\farcs$4 \citep[in the cross-scan 
direction;][]{iras88}.  To determine better positions for the sample, 
we searched SIMBAD\footnote{\url{simbad.u-strasbg.fr/simbad/}} for 
positions and cross-checked those by searching for sources within 
15\arcsec\ in the AKARI/Infrared Camera Point Source Catalog 
\citep[IRC PSC;][]{ish10} at 9 and 18~\mum.\footnote{\url{
https://doi.org/10.26131/IRSA181}}

AKARI has counterparts to 10,599 of the 11,238 sources in the extended
LRS Atlas (94\%).  Of the 639 missing sources, 125 are in the original 
LRS Atlas of 5425 sources (i.e., 2.3\% from the LRS Atlas are missing).
The remaining 564 missing sources represent 9.7\% of the 5813 
additional sources.  The two telescopes have similar sizes and 
resolutions.  The AKARI primary was 0.67~m in diameter, giving a 
resolution element at 12~\mum\ of 4$\farcs$6 compared to 5$\farcs$3 for 
IRAS.

Table~\ref{t.akarilrs} presents the results of the AKARI search by 
focusing on what is missing, broken down by the object type as 
defined by the LRS characterizations.  Table~\ref{t.akarivolk} does the 
same for the spectral classifications defined by \citet[][also referred 
to as the VC89 classes]{vol89a}.  These tables provide brief 
definitions of the classifications, while Appendix~\ref{s.vclass}
describes them and compares the two systems more thoroughly.  

AKARI recovers the vast majority of the stellar sources, and the
difference between the original atlas and the full extended atlas
is small.  Comparing LRS characterizations 1$n$, 2$n$, and 4$n$ to 
the similar VC89 classes, S, F, E, and C, gives similar rates of 
missing sources, 1.7\% and 2.1\%, respectively.

The percentage of missing sources rises for the redder sources.
For LRS characterizations 5$n$--9$n$, AKARI missed 8.9\% of 
the sources, compared to 20.3\% of VC89 classes A, P, H, and L.
The latter percentage includes sources in the original LRS
Atlas, plus the supplemental sources.  The redder sources tend to 
be associated with star-forming regions, which have complex 
backgrounds, are generally close to the Galactic plane and 
therefore in crowded fields, and can be extended.  For the
supplemental sources, all of those issues are more pronounced.
Both IRAS and AKARI would be challenged to identify targets
with these characteristics, so the higher percentage of missing
targets is expected, for both the red spectra in general and the 
supplemental red spectra in particular.  Plus, we were matching
to the AKARI point-source catalog, which selects against the 
more extended targets that can appear in the LRS database.  

The red spectra in the LRS database should be treated with some
caution.  Because the LRS data were obtained as IRAS scanned across 
the sky with no slit, it is possible that the wavelength calibration 
may not be as accurate for some extended sources.  Furthermore,
the scanning nature of the spectrometer leads to greater positional
uncertainties for the extended sources.  For sources in complex 
fields in the Galactic plane, more recent surveys should be
consulted to better determine what structures were likely to
contribute to the LRS spectra, such as the MSX survey \citep{pri01} 
and the GLIMPSE survey \citep{ben03, chu09}.\footnote{Galactic 
Legacy Infrared Midplane Survey Extraordinaire.}

\section{Summary \label{s.summary}} 

The new version of the extended LRS atlas contains 11,238 mid-infrared
spectra obtained by IRAS in 1983 covering nearly the entire
sky.  That total comprises all of the 5425 spectra in the original
LRS Atlas; 5796 from the additional data published by \cite{vol89a},
\cite{vol91}, and \cite{kwo97}, and 17 extra unpublished spectra
previously available on data servers.  The spectra have a spectral
resolving power between 20 and 60.

This version of the LRS atlas has corrected the spectra to remove the 
SiO artifact centered at 8~\mum\ present in earlier available 
databases.  However, the derivation of that correction with different
populations of stars reveals systematic differences that expose
some limitations in the spectra.  We estimate that the spectral
fidelity is $\sim$2--3\%.  Separate photometric analysis reveals
a $\sim$13\% difference between the calibration of the LRS on IRAS 
compared to the PSC for bright stars, and that difference increases by 
another $\sim$10\% for fainter stars in the LRS Atlas.

The sources added after the release of the original LRS Atlas greatly
improve the sampling close to the Galactic plane.  The vast majority of 
the spectra in the database arise from point sources, as indicated by 
the recovery of 94\% of them in the AKARI/IRC PSC.
These point sources are predominantly dust-free stars or stars with
optically thin dust shells.  Roughly 11\% of the sample are red
spectra, which may arise from extended sources and complex fields and
may have less reliable wavelength calibrations.

The extended LRS atlas remains the closest we have to a complete 
spectral survey of the mid-infrared sky.  The total number of sources, 
11,238, compares favorably to such fundamental surveys as the Yale 
Bright Star Catalog, which includes 9110 stars and defines the bright 
optical sky.  In the same way, the extended LRS atlas defines the 
bright mid-infrared sky.  It is now back in the public domain!

\begin{acknowledgements}

We thank the anonymous referee for helping us improve this manuscript.
G.~C.~S.\ and K.~E.~K.\ were supported by grant 80NSSC21K0985 from 
NASA's Astrophysics Data Analysis Program.
This work made use of NASA's Astrophysics Data System (funded by NASA 
under Cooperative Agreement 80NSSC21M00561), the NASA/IPAC Infrared 
Science Archive (funded by NASA and operated by the California 
Institute of Technology), observations with AKARI (a JAXA project with 
the participation of ESA), and the SIMBAD and VizieR databases 
(operated at the Centre de Donn\'{e}es astronomiques de Strasbourg).

\end{acknowledgements}

\facilities{IRAS,  ISO (SWS), Spitzer (IRS), AKARI (IRC)}

\appendix

\section{Defining the sample \label{s.defsample}} 

The sample of spectra from the LRS published here is based on the 
source list for the extended LRS catalog published by \cite{kwo97}
and the spectra in the Calgary database.  Not all targets from those
two sources made it to the final catalog. 

\cite{kwo97} published a list of 11,224 spectra, which included the 5425 
spectra from the original catalog \citep[the LRS Atlas;][]{lrs86} and 
5799 supplemental spectra.  That latter group included 386 spectra from 
\cite{vol89a}, 456 from \cite{vol91}, and 4957 new sources.  For the 
discussion below, we will refer to these three groups of spectra as the 
VC89, V91, and K97 sources.  We kept all but three of these spectra, 
omitting one from V91 and two from K97.  In addition, we renamed three of 
the sources.  Table~\ref{t.supp} summarizes the changes.

\begin{deluxetable}{lll} 
\tablecolumns{3}
\tablewidth{0pt}
\tablenum{5}
\tablecaption{Rejected or renamed sources from Kwok et al.\ (1997)}
\label{t.supp}
\tablehead{ \colhead{Target} & \colhead{Action} & \colhead{Reason} }
\startdata
02187$-$0302  & Rejected & Redundant; typo for 02168$-$0312 \\
05580$+$1633  & Rejected & Redundant; typo for 05583$+$1633 \\
07136$+$2851  & Renamed  & Now 07316$+$2851 \\
16262$-$2619A & Renamed  & Now 16262$-$2619 \\
19200$+$2101  & Rejected & Redundant with 19199$+$2100 \\
Egg Nebula    & Renamed  & Now 21003$+$3629M
\enddata
\end{deluxetable}

The rejected source from V91 is IRAS 19200+2101.  SIMBAD considers 
this source and two others (IRAS~19199+2100 and IRAS 19201+2101) to be a 
blend of two objects and gives all three the same coordinates.  SIMBAD 
reports that the actual objects are OH 55.1 +3.1, an OH/IR star, and 
UCAC4 556-089440, a star about which little is known.  The LRS spectrum of 
IRAS 19201+2101 shows silicate absorption at 10~\mum, which is consistent
with the OH/IR star.  The spectra of IRAS~19199+2100 and IRAS 19200+2101 
in the Calgary database are identical, and both are red, making them a 
poor match to a star.  The SIMBAD association to UCAC4~556-089440 appears
to be incorrect.  We kept IRAS~19199+2100 and dropped IRAS~19200+2101, 
but it was arbitrary which spectrum should be excluded.

Two K97 sources were also rejected.  IRAS~02187$-$0302 is not in the PSC 
and has a spectrum identical to the VC89 source IRAS~02168$-$0312 (better
known as Mira).  IRAS~05580+1633 is not in the PSC, has no LRS data 
associated with it in the database, and appears to be a typo for the K97 
source IRAS~05583+1633.

Three other supplemental sources were renamed.  IRAS~07136+2851 is a 
typo for IRAS~07316+2851 in the original LRS Atlas.  
Antares ($\alpha$ Sco) was identified as IRAS~16262$-$2619A, but in the
list by VC89 (and in the PSC) it is IRAS~16262$-$2619.  Finally, the 
Egg Nebula (AFGL~2688) is one of the three sources recovered from the 
strip in Cygnus with no counterparts in the PSC.  We assigned these 
three sources PSC-like names based on their 1950 coordinates and appended 
the suffix ``M'' for ``missing.''  Thus, AFGL~2688 becomes 
IRAS~21003+3628M.

\begin{deluxetable*}{lllc} 
\tablecolumns{4}
\tablewidth{0pt}
\tablenum{6}
\tablecaption{Extra sources}
\label{t.extra}
\tablehead{ \colhead{Target} & \colhead{Action} & \colhead{Reason} & 
            \colhead{Class}}
\startdata
01555+5239    & Kept     & \nodata                        & U \\
05437+2420    & Kept     & \nodata                        & E \\
05588$+$1633  & Rejected & Redundant, typo for 05583+1633 & \nodata \\
06230$+$1748  & Rejected & Redundant with 06230+1749      & \nodata \\
09553$-$6150  & Kept     & \nodata                        & F \\
16445$-$4459  & Kept     & \nodata                        & I \\
16458$-$4512  & Kept     & \nodata                        & H \\
16492$-$4349  & Kept     & \nodata                        & A \\
17006$-$4215  & Kept     & \nodata                        & H \\
17118$-$3909  & Kept     & \nodata                        & H \\
17143$-$3700  & Kept     & \nodata                        & A \\
17230$-$3459  & Kept     & \nodata                        & L \\
17478$-$2649  & Kept     & \nodata                        & A \\
V3811 Sgr     & Renamed  & Now 18206$-$2157               & H \\
18329$-$0629  & Kept     & \nodata                        & U \\
18358$-$0647  & Kept     & \nodata                        & A \\
18408$-$0353  & Kept     & \nodata                        & H \\
NML Cyg       & Renamed  & Now 20445+3955M                & F \\
20445$+$3956  & Rejected & Redundant with 20445+3955M     & \nodata \\
V407 Cyg      & Renamed  & Now 21004+4534M                & E \\
A05336$+$6846 & Rejected & Spectrum is just zeros        & \nodata \\
R Leo         & Rejected & Redundant with 09448+1139      & \nodata
\enddata
\end{deluxetable*}

Table~\ref{t.extra} lists the 22 extra unpublished sources in the
Calgary database.  We retained 17 of those, and of those 17, we 
renamed three because the Calgary server did not provide PSC-based names.
V3811~Sgr appears in the PSC as IRAS 18206$-$2157.  NML~Cyg and V407~Cyg 
are in the strip of Cygnus missing from the PSC.  As with AFGL~2688,
they do not have PSC counterparts, and we generated names appended
with ``M.''

Five of the extra spectra in the Calgary database were excluded.  The 
target identified as IRAS~A05336+6846 has a spectrum with just zeros for 
its signal, and no source close to this position appears in the PSC.  The 
remaining four rejects duplicate other spectra.  R~Leo appears twice, 
once with its variable-star designation and once as the VC89 source 
IRAS~09448+1139; we kept the latter.  Similarly, we have IRAS~06230+1748 
and the K97 source IRAS~06230+1749.  Both spectra are identical, and we 
retained the latter designation.  IRAS~05588+1633 was rejected because it 
appears to be another typo for IRAS~05583+1633 (a K97 source) and has no 
data associated with it.  
IRAS~20445+3956 was rejected because it is not in the PSC, and it has a 
spectrum identical to another extra source, NML Cyg, which we have 
already renamed to IRAS~20445+3955M.

For the surviving 17 extra sources, we classified them using the system
developed by \cite{vol89a} and described in Appendix~\ref{s.vclass}.
Two of the spectra had classifications on the Calgary server, and those 
remain unchanged:  ``F'' for IRAS~09553$-$6150, and ``I'' for 
IRAS~16445$-$4459.  A third source, IRAS~17230$-$3459,  had the undefined 
classification ``d'' on the Calgary server, and we have classified it 
as ``L'' due to the emission lines in its spectrum.

Thus, out of 11,224 spectra from K97, we have omitted three,
leaving 11,221.  Of the 22 extra spectra, we have kept 17, for a 
total of 11,238 spectra.

\section{Determining the spectral correction \label{s.correction}} 

For the red giants, the sample used for the spectral correction started
with the list of 13 bright red giants used as infrared standard stars in
Table~\ref{t.redgiants}.  Nine of those stars appear in Table~1 
from \cite{eng06}, which they describe as their ``best secondary
standards,'' after the two primary standards, the A dwarfs $\alpha$~CMa 
and $\alpha$~Lyr.  Of those nine red giants, we already rejected 
$\mu$~UMa (see Section~\ref{s.phot}).  For the remaining eight, we
use as a spectral template the Hanscom spectrum, which for seven of 
the eight is based on the spectrum from the SWS and shifted to be
consistent with the photometry.  The exception is $\beta$~Gem,
for which \cite{eng06} substituted spectra of stars with similar
spectral types, using the SWS spectrum of $\theta$ Cen (K0- IIIb) below 
9~\mum\ and the spectrum from ISOCAM on ISO of $\delta$ Dra (G9 III)
past 9~\mum.  For cooler K giants, the variation in strength of the SiO 
band within a spectral class could lead to errors in the strength of the
band, but a K0 giant shows little SiO absorption, so any errors should
be small.  

\begin{figure} 
\includegraphics[width=3.4in]{\figpath 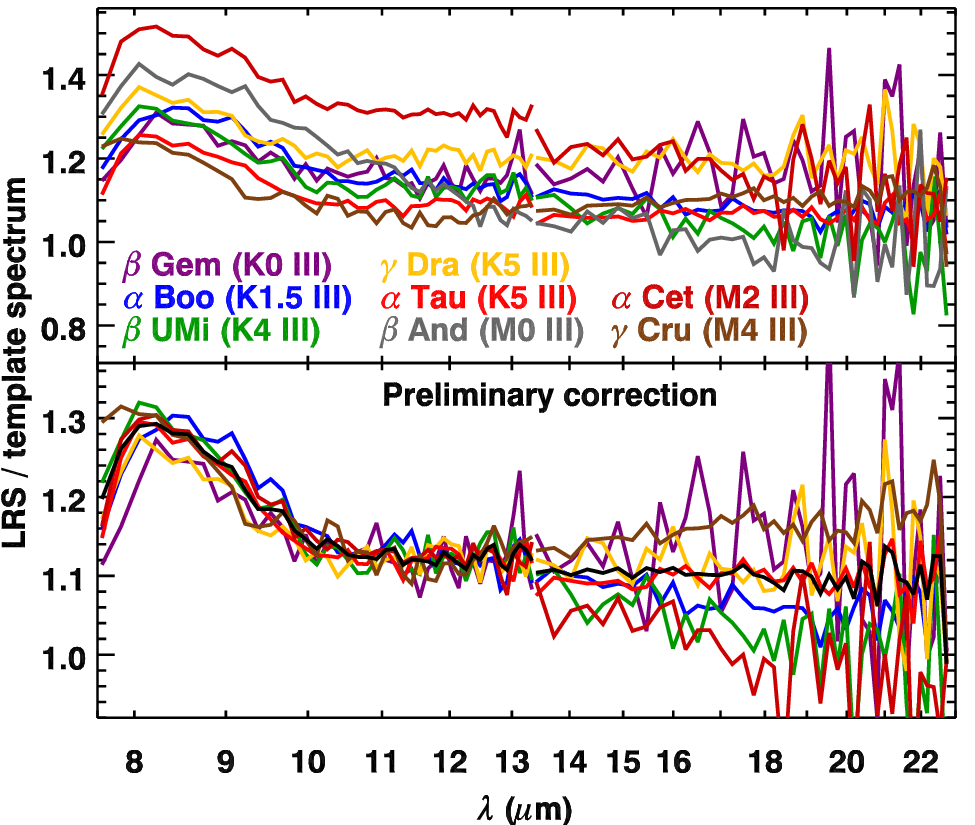}
\caption{Assembling the spectral correction for the red giants.  
\textit{Top:}  the ratios of the LRS spectrum divided by the template
spectrum for the same star.  
\textit{Bottom:}  the same ratios, after normalization to the 
flux-weighted mean at 10.6--13.4~\mum, along with the preliminary
spectral correction derived from the individual ratios (in black).
\label{f.corr_red} }
\end{figure}

\begin{figure} 
\includegraphics[width=3.4in]{\figpath 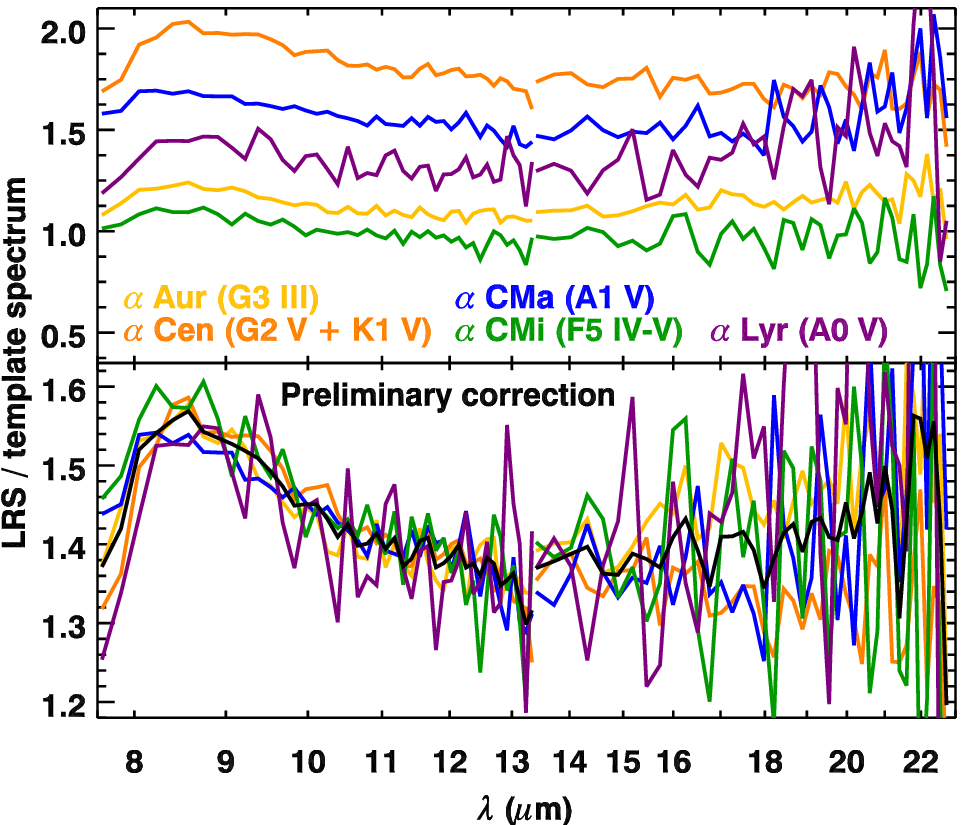}
\caption{Assembling the spectral correction for the warm stars.  As in
Figure~\ref{f.corr_red}, the top and bottom panels show the ratios
of LRS to spectral template before and after normalization at 
10.6--13.4~\mum.  The plotting range in the top panel is much wider
than in Figure~\ref{f.corr_red} to accommodate the greater spread in
the ratios of spectrum to template.  \label{f.corr_warm} }
\end{figure}

\begin{deluxetable}{lclr} 
\tablecolumns{4}        
\tablewidth{0pt}
\tablenum{7} 
\tablecaption{Bright warm stars used as infrared standards}
\label{t.warm}
\tablehead{ \colhead{ } & \colhead{IRAS} & \colhead{Spectral} & 
            \colhead{PSC} \\
            \colhead{Target} & \colhead{PSC} & \colhead{Type} &
            \colhead{$F_{12}$ (Jy)\tablenotemark{a}} }
\startdata
$\alpha$ Lyr & 18352$+$3844 & A0 V        &  28.7 \\
$\alpha$ CMa & 06429$-$1639 & A1 V        &  98.7 \\
$\alpha$ CMi & 07366$+$0520 & F5 IV--V    &  54.5 \\
$\alpha$ Cen & 14359$-$6037 & G2 V + K1 V & 153.2 \\
$\alpha$ Aur & 05130$+$4556 & G3 III      & 162.8 \\
\enddata
\tablenotetext{a}{Color-corrected by dividing by 1.45}
\end{deluxetable}

The top panel of Figure~\ref{f.corr_red} plots the ratio of the
uncorrected LRS spectrum to its template for the eight red giants.
Seven show a similar shape, with the SiO band apparent
(with a flipped sign) from the blue end to $\sim$10.6~\mum, and a
flat structure from there to the long-wavelength end of the blue
segment at 13.4~\mum.  The exception is $\beta$~And (in gray), which 
falls steadily past 10.6~\mum.  We have removed it from the sample.
The flux-weighted mean ratio from 10.6 to 13.4~\mum\ of the seven 
surviving spectra is 1.123 $\pm$ 0.025 (uncertainty in the mean), 
which compares favorably to the ratio of 1.129 $\pm$ 0.026 found in 
Section~\ref{s.phot} with a larger sample of red giants and a smaller 
wavelength range.  Much of the uncertainty in the sample considered 
here is driven by $\alpha$~Cet (ratio 1.306).  The large star-to-star 
variations reinforce the conclusion that the absolute photometric
calibration of the LRS is not reliable and further support the 
decision to not correct the LRS database photometrically.

The bottom panel of Figure~\ref{f.corr_red} shows the spectral ratios 
after scaling them to the same ratio between 10.6 and 13.4~\mum\ 
(1.123).  All seven ratios show similar structure through the SiO band.  
The mean spectral ratio with each star weighted by its PSC flux density 
gives the preliminary spectral correction based on the red giants.

As Figure~\ref{f.bright2} shows, the spectral correction based solely 
on red giants forces the spectrum of $\alpha$~CMa downward at the blue 
end (below $\sim$9.5~\mum) even as it improves the shape of the SiO band 
in $\alpha$~Tau compared to the Cohen correction.  Therefore, we also
generated a correction using five bright stars with higher effective 
temperatures than the red giants.  Table~\ref{t.warm} lists the sample.  
The A dwarfs $\alpha$~CMa and $\alpha$~Lyr are the current and former 
all-sky standards, respectively, and \cite{eng06} include 
$\alpha$~Cen and $\alpha$~Aur in their lists of secondary standards 
(their Tables 2 and 3, respectively).  For all four, we use Hanscom 
spectra as templates.  For the two A dwarfs, these templates are based 
on models.  For $\alpha$~Cen, the template is based on the SWS data, 
scaled to the photometry calibrated to align with the MSX calibration.  
For $\alpha$~Aur, the Hanscom spectrum is a template based on the 
star's spectral type and scaled to the photometry.  We excluded one 
warm star listed by \cite{eng06}, $\beta$~Dra, because it is too 
faint.  We added a fifth star, $\alpha$~CMi, because it is bright and 
easily characterized.   For its template, we used the BOSZ model 
generated as described in Section~\ref{s.processing}.

Figure~\ref{f.corr_warm} shows the construction of the preliminary 
spectral correction for the warm stars listed in Table~\ref{t.warm}.  
The top panel shows the spectral ratios before they have been 
normalized.  The most obvious difference between the warm stars and the 
red giants is the much wider range of spectral ratios.  The bottom
panel shows that despite that wide range, the ratios are reasonably
consistent in the wavelength range covered by the SiO artifact.

The mean correction from 10.6 to 13.4~\mum\ is based on a flux-weighted 
average of 1.375 with an uncertainty in the mean of 0.140.  The
spread in the ratios of LRS to template in the warm stars is 
significantly larger than in the red giants and reinforces the case
for not attempting a photometric correction to the LRS database.

Beyond 13.4~\mum, i.e., in the red spectral segment, the spectral
ratios vary significantly from star to star in both the red giants 
and the warm stars, even after normalization at 10.6--13.4~\mum.
A uniform spectral correction applied to all data would do nothing
to improve the situation, leading us to set the correction to 1.0
for all of the red spectral segment.

\begin{figure} 
\includegraphics[width=3.4in]{\figpath 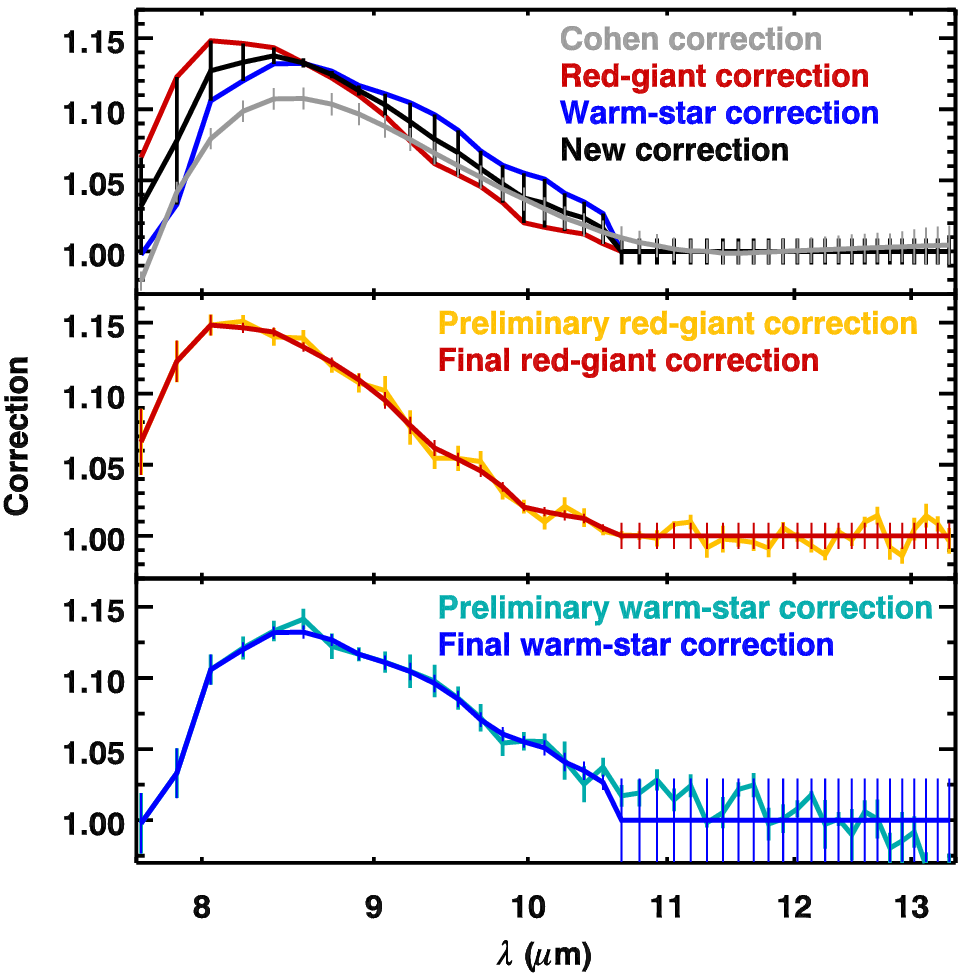}
\caption{Combining the spectral corrections from the red giants and 
warm stars.  \textit{Top:}  the new correction is the average of
the corrections for the red giants and warm stars determined in the
bottom two panels, and the Cohen correction is included for
comparison.  \textit{Middle and bottom}:  determining the final
corrections for the red giants and warm stars, respectively.   The
orange and light-blue curves give the preliminary corrections,
and the dark-red and dark-blue curves give the final corrections.
The error bars in the Cohen correction are as reported by 
\cite{coh92b}; all other error bars are the uncertainty in the mean.
\label{f.corr_final} }
\end{figure}

Figure~\ref{f.corr_final} shows how the new spectral correction has 
been determined from the preliminary corrections for the red giants 
and warm stars.  For both samples, the preliminary corrections 
in the bottom panels of Figures~\ref{f.corr_red} and \ref{f.corr_warm}
are normalized to 1.0 at 10.6--13.4~\mum\ by dividing by the mean
ratios, 1.123 and 1.375, respectively.  The spectra are then
smoothed with a 3-pixel boxcar to reduce the noise.  For the red 
giants, the spectra are flat from 10.6 to 13.4~\mum\ and normalized to 
1.0 there, making any spectral structure in that range likely just 
noise.  Therefore, at 10.6--13.4~\mum, we have set the correction to 
1.0 and the uncertainty in the correction to the standard deviation of 
the corrections at each wavelength, 0.0091.  The correction from the 
warm stars receives the same treatment, except that the uncertainty is 
0.0293.  These uncertainties in the correction are extended to the red 
end of the red segment.

The top panel of Figure~\ref{f.corr_final} shows the new correction,
which is the mean of the corrections from the two samples.  The 
uncertainties redward of 10.6~\mum\ are taken directly from the 
red-giant correction.  The corrections from the red giants and
warm stars for the SiO artifact differ systematically, with the
warm-star correction slightly weaker and shifted to longer 
wavelengths.  It also drops back to $\sim$1.0 at the blue end of
the spectrum.  These differences are a concern, because all of the
spectra in the LRS database should show identical artifacts and
require the same spectral correction.  No obvious cause for the
differences between the warm stars and red giants stands out.
Generally, the red giants are brighter than the warm stars, but
within the red giants, Figure~\ref{f.corr_red} does not reveal
any significant differences between the three brightest red giants,
$\gamma$~Cru, $\alpha$~Boo, and $\alpha$~Tau, all with $F_{12}$ 
$\sim$ 500--600~Jy, and the rest of the sample.  Similarly, the
three faintest red giants, $\beta$~Gem, $\beta$~UMi, and 
$\gamma$~Dra, with $F_{12}$ $\sim$ 80--110~Jy, do not stand out
as a group.  Two have slightly weaker artifacts to correct for,
but otherwise their shape resembles the rest of the red giants.

Figure~\ref{f.corr_final} includes the Cohen correction for
comparison.  It resembles the warm-star correction somewhat more 
than the red-giant correction, although it is weaker, with a peak
correction $\sim$10\%, versus 13\% for the warm stars and
14\% in the combined new correction.

\section{Infrared spectral classifications \label{s.class}} 

\cite{kwo97} provided infrared spectral classifications based on the 
VC89 scheme for the entire sample using simple one-letter designations.  
The 5,425 spectra in the original LRS Atlas also have a two-digit LRS
characterization that provides more information \citep{lrs86}.
We have retained both of these classifications, and this 
appendix describes and compares them.  It also compares the VC89
classifications to schemes developed for the spectrometers aboard ISO.
Tables~\ref{t.lrschar} and \ref{t.vclass} outline the LRS 
characterizations and the VC89 system, and Table~\ref{t.vclass} 
gives approximate equivalences between the two systems.

\begin{deluxetable*}{llrll} 
\tablecolumns{5}
\tablewidth{0pt}
\tablenum{8}
\tablecaption{LRS Characterizations}
\label{t.lrschar}
\tablehead{ \colhead{LRS} & \colhead{ } & \colhead{ } &
  \colhead{Fraction of} & \colhead{Second} \\
  \colhead{Char.} & \colhead{Description} & \colhead{N} & 
  \colhead{Total (\%)} & \colhead{Digit} }
\startdata
0$n$ & Unusual or low S/N spectra        &  363 &  6.7 & 
       $n$ = 0--5, depending on the spectral shape \\
1$n$ & Blue featureless spectra          & 2238 & 41.3 &
       $n$ = $-$2 $\beta$, where $F_{\lambda} \sim \lambda^{\beta}$ \\
2$n$ & Silicate emission at 10~\mum      & 1730 & 31.9 &
       $n$ increases with feature strength\\
3$n$ & Silicate absorption at 10~\mum    &  230 &  4.2 & $n$ as 2$n$ \\
4$n$ & SiC dust emission at 11.3~\mum    &  538 &  9.9 & $n$ as 2$n$ \\
5$n$ & Red featureless spectra           &   63 &  1.2 & 
       $n$ = 2 $\beta$, where $F_{\lambda} \sim \lambda^{\beta}$ \\
6$n$ & As 2$n$, but with a red continuum &   78 &  1.4 & 
       $n$ increases with feature strength\\
7$n$ & As 3$n$, but with a red continuum &   67 &  1.2 &  $n$ as 6$n$ \\
8$n$ & PAHs and emission lines           &   69 &  1.3 &
  $n$ based on the strongest line \\
9$n$ & Emission lines and no PAHs        &   49 &  0.9 & $n$ as 8$n$
\enddata
\end{deluxetable*}

\begin{deluxetable*}{llrrl} 
\tablecolumns{5}
\tablewidth{0pt}
\tablenum{9}
\tablecaption{Spectral classes from Volk \& Cohen (1989) (VC89)}
\label{t.vclass}
\tablehead{ \colhead{Spectral} & \colhead{ } & \colhead{ } &
   \colhead{Fraction of} & \colhead{Analogous LRS} \\
   \colhead{Class} & \colhead{Description} & \colhead{N} & 
   \colhead{Total (\%)} & \colhead{Characterizations\tablenotemark{a}} }
\startdata
S & Dust-free stellar continuum            & 1635 & 14.5 &
    \textbf{16--19}, 01 \\
F & Featureless nonstellar blue continuum  & 1944 & 17.3 &
    \textbf{13--16}, 01 \\
E & Silicate emission at 10~\mum           & 3617 & 32.2 &
    \textbf{2$n$}, 13--16, 42--43, 69, 01 \\
A & Silicate absorption at 10~\mum         &  319 &  2.8 & \textbf{3$n$} \\
C & SiC dust emission at 11.3~\mum         &  715 &  6.4 & 
    \textbf{4$n$}, 14--17 \\
P & PAH emission features                  &  315 &  2.8 &
    \textbf{80}, 32--35, 81 \\
H & Red continuum                          &  638 &  5.7 &
    \textbf{91}, 80--81, 05, 7$n$, 5$n$ \\
L & Emission lines                         &   31 &  0.3 & 94--95 \\
U & Unusual \textbf{spectrum}              &  661 &  5.9 &
    12--15, 50, 21--23, 01 \\
I & Incomplete or noisy \textbf{spectrum}  & 1363 & 12.1 & 01, 05, 1$n$, 50
\enddata
\tablenotetext{a}{The predominant LRS characterizations are in bold.}
\end{deluxetable*}

\subsection{LRS characterizations \label{s.lrschar}} 

The LRS characterization scheme was partially automated, starting by
fitting a power law to the spectra from 14 to 22~\mum.  If the spectral 
index $\beta$ was less than $-$1, where $F_{\lambda} \sim \lambda^{\beta}$, 
then the spectrum was assigned to 1$n$ through 4$n$, depending on the 
dominant spectral features.  Otherwise, the spectrum was assigned to 
5$n$--9$n$.  Noisy spectra or spectra that did not fit the other 
categories were assigned to 0$n$.  Table~\ref{t.lrschar} simplifies the 
descriptions of the subclasses ($n$) given by Table 1 from \cite{lrs86}, 
which should be consulted for further details.

Blue spectra dominated the LRS Atlas, with the three most populous classes 
being the featureless spectra (1$n$), the silicate emission spectra (2$n$), 
and the carbon-rich spectra (4$n$).  Combined, these spectra account for 
83\% of the original LRS Atlas.  Dust-free stars generally had LRS 
characterizations of 18, since the Rayleigh-Jeans tail of a star should 
fall as $\beta = -4$ (and $n = -2 \beta$), with noise spreading the range 
of characterizations to 17--19.  Sources with low-contrast dust emission 
(typically from alumina) fell into the range 14--16 \citep{slo95}.  Sources
with low-contrast oxygen-rich or carbon-rich dust emission can wind up
with characterizations anywhere in the low 20s, 30s, or 40s because of 
noise in their spectra.  This confusion can be seen in tables with
LRS characterizations of independently verified dust chemistries, such
as oxygen-rich dust in S stars \citep[][their Table 1]{lml88}, dust in
oxygen-rich AGB stars \citep[][their Table 4]{slo98b}, and dust in carbon
stars \citep[][their Table~2]{slo98a}.

These inconsistencies in relating the LRS characterizations directly
to classes of objects (dust-free stars, oxygen-rich stars with 
optically thin dust shells, etc.) arise from the limited S/N in the
sample.  This problem grows with the fainter samples that dominate
the sources added in the supplemental catalogs, making it unwise
to apply the LRS characterizations to the full sample of 11,238
spectra.  Each spectrum does have a classification in the system
introduced by \cite{vol89a}, making it the better system for 
navigating the full LRS database.

\subsection{VC89 classifications \label{s.vclass}} 

The ``S'' class is for stellar spectra with no dust, while the 
``F'' class includes other featureless blue spectra that are 
falling less steeply with wavelength than a Rayleigh-Jeans tail, most 
likely because of low-contrast dust emission.  Effectively, these two 
classes split the 1$n$ LRS characterizations.  Of the 1171 
class S spectra in the original LRS Atlas, 751 (64\%) are 17 
or 18, with another 161 characterized as 01 and about 100 
each as 16 and 19.  Of the 1014 class F spectra with LRS 
characterizations, 759 (75\%) are in the range 14--16, 90 
are 01, and 54 are class 13.

Most optically thin oxygen-rich dust shells are classified as ``E'' 
to denote emission from silicate and related dust.  Of the 
2082 in the LRS Atlas, 1659 (80\%) have LRS characterizations of 
2$n$.  Another 250 (12\%) are in the range 12--17, with most in 13--16.  
The confusion of low-contrast carbon-rich and oxygen-rich dust emission 
has placed 55 in the range 41--45 (mostly 41--43).  Interestingly, 45 of 
the class E sources have LRS characterizations of 69, 
indicating strong silicate emission and a red continuum.  These sources 
are primarily oxygen-rich post-AGB objects.  Another 37 are in the 0$n$ 
range, most of them 02.

The optically thick oxygen-rich dust shells with silicate absorption
features at 10~\mum\ and/or 18~\mum\ are classified as ``A'' and are 
better behaved than the class E sources.  Of the 125 in the LRS 
Atlas, 115 (92\%) are in the 3$n$ sequence.

The carbon-rich ``C'' class includes 565 sources in the LRS 
Atlas, with most of these (437, or 77\%) on the 4$n$ sequence.  Of the 
remainder, 87 (15\%) are characterized as 1$n$ (most of them 14--17), 21 
are characterized as 0$n$, and 16 are scattered among 21--24 and 31--32.

The remaining VC89 classes are less populated in the LRS Atlas 
and tend to be scattered across more characterizations, making them 
more difficult to map onto the LRS characterizations.  The PAH emission
sources (``P'') are a good example.  Of the 121 in the LRS Atlas, 27 
have an LRS characterization of 80 (PAHs and no emission lines), 
with 9 others on the 8$n$ sequence, 47 on the 3$n$ sequence, 
15 on the 1$n$ sequence, and 10 on the 7$n$ sequence.  The ``U'' class
catches many unusual spectra.  Many have flat continua or cooler 
blackbody spectra that peak in the LRS wavelength range, which implies 
a temperature of 200--300~K.  For a spectrum to be classified as ``H''
(for \ion{H}{2} regions), it must continue rising to the 
long-wavelength end of the spectrum.  Classes ``L'' and ``I'' include
spectra dominated by emission lines and incomplete spectra, respectively.

While the LRS characterizations and the VC89 classes are most relevant
to the new extended LRS atlas, other classification systems are also
available.   A Bayesian classification scheme led to what has become 
known as the Autoclasses \citep{che89}, which placed the 5425 spectra 
in the original atlas in 77 self-defined classes.  \cite{lml88, lml90} 
classified over 400 optically thin oxygen-rich circumstellar dust 
shells in the original atlas by the shape of the spectral emission 
profile instead of the strength of the feature that drove the LRS 
characterizations.  \cite{slo95, slo98b} generalized the shape-based
approach by defining a silicate dust sequence and applied it to over 
600 LRS spectra.  That number is still just a fraction of the total of
oxygen-rich dust sources in the extended LRS atlas.  

\subsection{Comparison to ISO-based classification systems} 

Comparing the VC89 system to the classification scheme developed by
\cite{kra02} for the ISO/SWS reveals both its strengths and its
weaknesses.  The primary weakness of the VC89 system compared to the
SWS classication results from the limitations of the LRS data, with 
their lower spectral resolution, reduced wavelength coverage, and
lower angular resolution.  The full-scan spectra from the SWS cover
2.4--45~\mum\ with a spectral resolving power of several hundred,
compared to 7.67--23.73~\mum\ and 20--60 for the LRS.  These
improvements allowed a more thorough scheme for the SWS, which 
classified spectra in five groups from blue to red, with dust-free
stellar spectra in group 1, stars with increasing amounts of 
circumstellar dust in groups 2 and 3, and increasingly red spectra
from more evolved objects, nebulae, and young stellar objects in
groups 4 and 5.  Group 6 was reserved for spectra with no continuum,
and group 7 was reserved for spectra with major flaws in the data.  
The SWS classifications added descriptors to the group numbers such 
as ``N'' for naked (i.e., dust-free) stars, ``SE'' for silicate and
related oxygen-rich dust emission, ``SA'' for silicate dust 
absorption, ``CE'' for carbon-rich dust emission, and ``U'' for
PAH emission (previously known as the UIR, or unidentified infrared 
emission features).  This system leads to many more classifications
than just the 10 classes in the VC89 system.

\cite{kra02} found that the classifications of the stellar spectra in 
the VC89 system aligned closely with the SWS classifications.  We
repeated their analysis and found a total of 776 spectra of sources
in common between the SWS sample \citep[as defined by][]{slo03} after
excluding offset targets and group 7 spectra.  Of the 111 matched 
class S spectra, 90\% are classified as dust-free stars (1.N, with 
various subclasses), and most of the remainder are stars with 
low-contrast alumina dust emission (2.SEa).  Of the 172 matched
class E spectra, 65\% have SWS spectra classified as optically
thin oxygen-rich dust emission (2.SE), 18\% have optically thick 
silicate dust in emission or partial self-absorption (3.SE and 3.SB),
and another 7\% have silicate emission superimposed on a red
continuum (4.SE).  Of the 67 matched class C spectra, all but two
of the SWS spectra are classified as carbon-rich (49 in group 2;
16 in group 3).  The matched class F spectra align well with
low-contrast dust emission, with 55\% of the spectra classified as
2.SEa.  Folding in the 1.N and remaining 2.SE sources increases the 
percentage to 67\%, but the remainder are scattered, with 13\% in 
group 3 and 3\% in group 4.  To summarize, the spectra from the SWS
validate the majority of the VC89 classifications for the relatively
blue spectra in classes S, F, E, and C.

The redder VC89 classes, however, show substantially more scatter.
The best behaved group is class P.  Of the 45 matched spectra,
71\% have a ``U'' or ``u'' in their SWS classifications, with the
lowercase ``u'' indicating the presence of PAHs in a spectrum
dominated by something else.  Another 20\% of the class P spectra
are matched to SWS spectra with silicate absorption in their
spectra, because both can produce similar shapes in the limited
wavelength range available to the LRS.  Of the 162 matched class H
spectra, nearly all are in groups 4 or 5, indicating that the SWS
spectra confirms their red continua.  However, 31\% are 
associated with SWS class 5.UE, indicating that PAH emission features
rival the atomic emission lines in strength.  It follows that users 
searching for PAH emission should not limit themselves to the P 
class.  The SWS classifications for the 25 matched L spectra reveal 
that 92\% of them have continuum shapes and spectral features 
suggesting that they are planetary nebulae.

The remaining classes, U and I, are more difficult to generalize,
which should be expected for unusual and incomplete spectra.  The
76 matched class U spectra, interestingly, include 25 reddened 
carbon-rich spectra (with classifications 4.CR, 4.CT, 4.CN, and 
4.C/SC).  Otherwise, the class U spectra are scattered from group 1 
to group 6.  The 11 matched class I spectra are predominately in 
group 5, probably because the sources are extended and in complex 
backgrounds where a complete spectrum could not be extracted.

\cite{hod04} also compared the VC89 classifications with the scheme 
developed by \cite{kra02}.  They classified spectra observed by the 
PHT-S spectrometer on ISO that cover the wavelength ranges 2.5--4.9~\mum\
and 5.8--11.6~\mum.  The PHT-S sample included 237 sources with spectra 
from the LRS, and they verified that most of the VC89 classifications 
of blue spectra agreed well between the LRS and PHT-S.  However, the 
differences in wavelength coverage and sensitivity between the two sets 
of spectra can occasionally lead to significantly different 
classifications.  From the current and previous comparisons, the lesson 
emerges that the VC89 classifications should be used as a guide and not 
relied on to be precise, more so for the redder classes than the bluer
ones.

\section{Accessing the extended atlas \label{s.access}} 

The individual spectra in the extended LRS atlas will be available 
from the Infrared Science Archive (IRSA) at 
IPAC.\footnote{\url{http://irsa.caltech.edu}}
The data are also available from Dataverse.\footnote{
\url{https://doi.org/10.7910/DVN/3028UI}}
Additional data files are also available, including the LRS
wavelength grid with the resolution at each wavelength element and 
the four spectral corrections on that grid described in this paper 
(the Cohen correction, the corrections from the red giants and warm 
stars, and the final correction from the combination of those 
two).\footnote{All data files are also available at\\
\url{https://users.physics.unc.edu/~gcsloan/library/lrsatlas}.}

\begin{deluxetable*}{lrrllrcrrrr} 
\tablecolumns{11}
\tablewidth{0pt}
\tablenum{10}
\tablecaption{The catalog of sources in the extended LRS atlas}
\label{t.catalog}
\tablehead{ 
  \colhead{Target} & \colhead{RA} & \colhead{Dec.} & \colhead{Position} & 
  \colhead{ } & \colhead{LRS} & \colhead{VC89} & 
  \colhead{IRAS} & \colhead{IRAS} & \colhead{AKARI} & \colhead{AKARI} \\
  \colhead{IRAS PSC} & \colhead{(deg, J2000)} & \colhead{(deg, J2000)} & 
  \colhead{Reference\tablenotemark{a}} & \colhead{Source} & \colhead{Char.} & 
  \colhead{Class} & \colhead{$F_{\nu}$ (Jy)\tablenotemark{b}} &
  \colhead{$F_{\nu}$ (Jy)\tablenotemark{b}} & 
  \colhead{$F_{\nu}$ (Jy)\tablenotemark{b}} & 
  \colhead{$F_{\nu}$ (Jy)\tablenotemark{b}} }
\startdata
00001$+$4826 & 0.684258 &    48.714115 & 2020yCat.1350....0G  & original & 
    21 & E & 48.870 & 24.100 &  33.78 & 14.15 \\
00007$+$5524 & 0.839422 &    55.681120 & 2020yCat.1350....0G  & original &
    22 & E & 97.600 & 46.920 & 124.00 & 58.51 \\
00012$+$6626 & 0.965171 &    66.712183 & 2020yCat.1350....0G  & original &
    18 & S & 12.690 &  3.220 &  17.22 &  3.97 \\
00017$+$3949 & 1.083645 &    40.109952 & 2020yCat.1350....0G  & original &
    16 & F & 17.740 &  6.850 &  21.39 &  8.92 \\
00019$+$4150 & 1.126582 &    42.119910 & 2020yCat.1350....0G  & kwok97   &
  $-$1 & F & 18.680 &  6.905 &  21.05 &  7.44 \\
00019$-$1047 & 1.125493 & $-$10.509524 & 2007A\&A...474..653V & original &
     1 & S & 12.730 &  3.363 &  17.07 &  3.95 \\
00020$+$4316 & 1.151698 &    43.551313 & 2020yCat.1350....0G  & original &
    18 & S & 10.420 &  3.385 &  13.19 &  3.87 \\
00036$+$6117 & 1.571383 &    61.565894 & 2003yCat.2246....0C  & kwok97   &
  $-$1 & F &  7.225 &  4.174 &  11.14 &  5.65 \\
00036$+$6947 & 1.559355 &    70.067281 & 2020yCat.1350....0G  & original &
    44 & C & 36.570 &  9.936 &  31.18 &  8.43 \\
00039$+$2648 & 1.621968 &    27.090110 & 2020yCat.1350....0G  & original &
    16 & F & 16.920 &  6.250 &  18.80 &  6.74
\enddata
\tablecomments{This table is available in its entirety in machine-readable 
  format.}
\tablenotemark{a}{The bibcode references are to (in order of mention)
  Gaia EDR3 \citep[released in 2020;][]{gaia21}, Hipparcos \citep{vl07}, 
  and the Two Micron All Sky Survey \citep[released in 2003;][]{skr06}.}
\tablenotemark{b}{The photometry is not color-corrected.}
\end{deluxetable*}

Each spectral data file has five columns:  wavelength (\mum), flux 
density (Jy), uncertainty in flux density (Jy), a segment identifier, 
and the raw flux density (Jy).  The flux density column is after the 
application of the spectral correction, and the raw flux density is 
before.  The uncertainties are all zeros because they are not 
provided in the original data, but they are included because some 
software will expect that column in the data files.  The segment 
identifier is 1 for the blue segment (7.67--13.34~\mum) and 2 for the 
red segment (13.41--22.73~\mum).  Each file has a header with the 
ancillary data in Table~\ref{t.catalog}.

For each spectrum, Table~\ref{t.catalog} provides the IRAS PSC name 
(reconstructed with ``M'' for the three missing sources), the position 
from SIMBAD (in degrees), and a reference for that position provided 
as a bibcode as referenced by SIMBAD and the Astrophysics Data 
System.\footnote{\url{http://ui.adsabs.harvard.edu}}  The source 
column has a value of ``original'' for the original LRS Atlas, 
``vc89'' for \citet[][V89]{vol89a}, ``volk91'' for \citet[][V91]{vol91}, 
and ``kwok97'' for \citet[][K97]{kwo97}.  The LRS characterizations 
have a value of $-$1 for sources not in the original LRS Atlas, and the 
VC89 classifications are as given by \cite{kwo97} or Table~\ref{t.extra}.  
Table~\ref{t.catalog} also has the uncorrected flux densities at 12 
and 25~\mum\ from the IRAS PSC and at 9 and 18~\mum\ from the 
AKARI/IRC PSC, all in Jy.  Missing fluxes have a value of $-$1~Jy.

\end{document}